\documentclass[authoryear,review,3p,times]{elsarticle}

 \usepackage{graphicx}

\usepackage{amsmath, amsthm, amssymb}
\usepackage{natbib}
\usepackage{lineno}
\journal{Icarus}

\begin{document}

\begin{frontmatter}



\title{The carbon-14 spike in the 8th century was not caused by a cometary impact on Earth}


\author[1]{Ilya G. Usoskin\corref{cor}}
\author[2]{Gennady A. Kovaltsov}

\address[1]{ReSoLVE Center of Excellence and Sodankyl\"a Geophysical Observatory (Oulu unit) University of Oulu, Finland}
\address[2]{Ioffe Physical-Technical Institute, St.Petersburg, Russia}
\cortext[cor]{Corresponding author, e-mail: ilya.usoskin@oulu.fi}


\begin{abstract}
A mysterious increase of radiocarbon $^{14}$C ca. 775 AD in the Earth's atmosphere has been recently
 found by Miyake et al. (Nature, 486, 240, 2012).
A possible source of this event has been discussed widely, the most likely being an extreme solar energetic particle event.
A new exotic hypothesis has been presented recently by Liu et al. (Sci. Rep., 4, 3728, 2014) who proposed that the event was caused by
 a {\bf cometary impact on Earth bringing additional $^{14}$C to the atmosphere}.
{\bf Here we calculated a realistic mass and size of such a comet to show that it would have been
 huge ($\approx 100$ km across and $10^{17}-10^{20}$ gram of mass) and would have produced a disastrous geological/biological impact on Earth.}
The absence of an evidence for such a dramatic event makes this hypothesis invalid.
\end{abstract}



\end{frontmatter}

\linenumbers

\section{Introduction}

Radiocarbon $^{14}$C is a primary cosmogenic radioisotope produced in the Earth's atmosphere by galactic cosmic rays
 whose flux is a subject to solar and geomagnetic modulation.
After production, radiocarbon is redistributed in the complicated terrestrial carbon cycle \citep{roth13} and finally stored
 in a natural archive like tree trunks or corals, where it can be measured later to reconstruct the cosmic ray
 flux and respectively solar activity in the past \citep{beer12,usoskin_LR_13}.
Occasionally however, sources other than galactic cosmic rays may contribute to the measured $^{14}$C.
One such known event is related to a strong increase of $^{14}$C around year 774-775 AD, discovered first in
 a Japanese cedar tree rings \citep{miyake12} and later confirmed by measurements in a German oak tree \citep{usoskin_775_13}.
A similar increase has been observed also in another cosmogenic radioisotope $^{10}$Be measured in Dom Fuji Antarctic ice
 core \citep{horiuchi08,usoskin_ApJ_12}.
Different possible sources of this event have been discussed in the literature, from a gamma-ray burst \citep{hambaryan13,pavlov13} to a
 cometary impact on the Sun \citep{eichler12}.
The most {\bf realistic source is} an extreme flux of solar energetic particles accelerated in
 a giant solar flare or interplanetary shock \citep{melott12,usoskin_775_13,cliver14}.

\citet{liu14} provided a new measurements of $^{14}$C in coral skeletons of the South China Sea for the period of 770s AD.
The measurements show a clear increase of $\Delta^{14}$C occurred in 773 AD as resolved with a great temporal resolution.
This confirms the earlier discovery of a 12 permill increase of $\Delta^{14}$C in tree rings in
 Japan \citep{miyake12} and Europe \citep{usoskin_775_13} for 774-775 AD.
In addition, \citet{liu14} reported a chronicle record of a {\bf meteor/comet impact on the Earth's atmosphere:
``It is well established that a comet collided with the Earth's atmosphere ... on 17 January AD 773''.}
Accordingly, they proposed that this comet might have brought additional $^{14}$C and $^{10}$Be into the atmosphere
 which would explain the observed peak.
This is a strong conclusion with potentially great scientific impact and thus requires a robust quantitative support.
However, the estimates of the comet's body offered by \citet{liu14} are too much uncertain.
Here we revisited this idea and estimated the realistic size and mass of a comet needed to produce the observed increase in $^{14}$C.
{\bf We show that an impact of such a comet would had been disastrous, but such an event did not take place,
 making the idea of the cometary origin of the measured $^{14}$C spike invalid.}

\section{Result and Discussion}

All the measurements in tree rings \citep{miyake12,usoskin_775_13} and in corals \citep{liu14} agree that the
 relative increase of $^{14}$C was 11--12 permill (1.1--1.2\%), and considering the complementing information of $^{10}$Be
 deposition in Antarctic ice \citep{horiuchi08}, the effect was global.
Such an increase of $\Delta^{14}$C requires an additional source {\bf (denoted here as $S$)} of $\approx 1.5\times 10^8$ $^{14}$C atoms per cm$^2$ in the atmosphere
 \citep{usoskin_775_13,pavlov13}.
By applying the global distribution, one can easily obtain that
 $N=S\times 4\pi\,R_{\rm Earth}^2=8\times 10^{26}$ atoms (or about 18 kg) of $^{14}$C should be instantly injected into the atmosphere.
Since the amount of radiocarbon in the comet's body is a result of a balance between production $Q$ and radioactive
 decay (the life-time $\tau=8266$ years) of $^{14}$C, $N=\tau\cdot Q$, one can calculate that $Q=3\times 10^{15}$ atoms of
 $^{14}$C should be produced in the comet's body per second during its transport in space.

Radiocarbon is mostly produced from nitrogen by capture of a {\bf secondary} (epi)thermal neutron $^{14}$N(n,p)$^{14}$C.
Other channels have much lower cross-sections \citep{kovaltsov12}.
Thus, the amount of nitrogen atoms in a thick target irradiated by cosmic rays is essential for its ability to produce $^{14}$C.
In this sense, the Earth's atmosphere containing 78\% of nitrogen is close to the absolutely effective target {\bf in a sense that
 production of $^{14}$C is the most effective sink of thermalized neutrons.}
According to numerical radiocarbon production models \citep{masarik09,kovaltsov12} the average $^{14}$C production rate in the Earth's atmosphere as
 produced by galactic cosmic rays in the absence of solar and geomagnetic shielding (as corresponding to an outer part of the solar system) is
 about 10 atoms/cm$^2$/s.
Thus, in order to guarantee sufficient production of $^{14}$C in the comet's body, its surface area (the body is assumed to be thick enough
 for the development of a cosmic-ray induced cascade to produce (epi)thermal neutrons) must be $3\times 10^{14}$ cm$^2$.
This corresponds to the mean radius of about $5\times 10^6$ cm or 50 km (100 km across) {\bf for the spherical shape}.
We note that $^{14}$C is produced only in an upper layer of tens of meters thickness, while the bulk of the volume is unreachable for cosmic rays.
With the mean comet density \citep{britt06} of about 0.6 g/cm$^3$, it leads to the total mass of a $3\times 10^{20}$ g.
A similar estimate can be found independently in \citet{melott13} based on direct numerical simulations.
This would make such a comet the biggest one observed.
A body of such size and mass would have produced a dramatic impact if falling on Earth.
A simple estimate of such a body at {\bf 30 km/s} velocity leads to the impact energy of $10^{26}$ J or $3\times 10^{10}$ megaton TNT
 {\bf which is a factor of $10^{9}$--$10^{10}$ greater than that for the Tunguska phenomenon \citep{borovicka13}.}
Such an impact would lead to dramatic geological/biospherical consequences that could not remain unnoticed.
{\bf If the comet was not a perfectly spherical body, hollow inside or even being a swarm of smaller objects, the estimated mass could be
 somewhat smaller.
An unrealistically conservative lower limit would be $3\times 10^{17}$ g, which would correspond to a swarm of about $10^7$ objects of 20 m across
 arranged so they do not shield each other of cosmic rays, or a thin sphere of 100 km across and 10 m thick.}

We note that all the assumptions made here {\bf tend to underestimate the comet's size and mass}.
For example, any composition other than considered here, of the comet would only lead to a less effective production of $^{14}$C and
 accordingly to the bigger size/mass required.
Thus, the present estimate is a conservative lower bound.


Thus, the hypothesis of a cometary impact bringing additional radiocarbon to the atmosphere in ca. 773 AD \citep{liu14} is proven invalid,
 and the solar origin of the event \citep{usoskin_775_13,cliver14} remains the most plausible explanation.

\section*{Acknowledgements}
G.K. was partly supported by the Program No. 22 presidium RAS and by the Academy of Finland.
Support by the Academy of Finland to the ReSoLVE Center of Excellence (project no. 272157) is acknowledged.


%
\end{document}